\documentclass[aps,prl,twocolumn,showpacs,amsmath,amssymb]{revtex4}
\usepackage{graphicx}
\def\Tr{\mbox{tr}}
\begin{document}
\title{Mesoscopic versus Macroscopic division of current fluctuations.}
\date{\today}
\author{V. Rychkov}
\email{rytchkov@physics.unige.ch}
\author{M. B\"uttiker}
\affiliation{D\'epartement de Physique Th\'eorique, Universit\'e
de Gen\`eve, CH-1211 Gen\`eve 4, Switzerland} 
\begin{abstract}
We investigate the current shot noise at a three terminal
node in which one of the branches contains a noise generating
source and the correlations are measured between the currents
flowing through the other two branches. Interestingly,
if the node is macroscopic, the current correlations are positive,
whereas for a quantum coherent mesoscopic node anti-bunching of
electrons leads to negative correlations. We present specific
predictions which permit the experimental investigation of the
crossover from quantum mechanical noise division to macroscopic
noise noise division.
\end{abstract}
 \maketitle
{\it Introduction --} Over the past two decades the theoretical
and experimental investigation of the noise properties of small
conductors has developed into a major field of research in
mesoscopic physics. Fundamentally shot noise is a consequence of
the granularity of charge and quantum diffraction \cite{review}.
In quantum coherent conductors shot noise arises whenever there
are multiple final states for a given incident state. For purely
elastic scattering, for conductors embedded in a zero-impedance
external circuit, the Pauli exclusion principle leads to negative
current correlations independent of the shape and form of the
conductor \cite{mb91}. Negative current correlations have been
measured at beam splitters and in quantum Hall effect geometries
\cite{henny,oliver,ober1}. If the occupation of the incident channel
is small and approaches a Maxwell-Boltzmann distribution, the
current correlations vanish \cite{ober1}.

In contrast, positive correlations are known to occur due to
interaction in superconducting-normal hybrid structures \cite{ns},
in  ferromagnetic spin controlled hybrid structures
\cite{spin,bruder}, or in normal conductors due to dynamical
screening \cite{BNazarov,martin}. However, recently Texier and
B\"uttiker \cite{texier} predicted positive correlations in the
white noise limit of purely normal conductors for a geometry
\cite{ober1} in which edge states are coupled to a voltage probe.
The correlations change sign as the coupling to the voltage probe
increases. A recent experiment by Oberholzer \emph{et al.}
\cite{ober2} is in excellent agreement with theory. Wu and Yip
\cite{yip} investigate the sign of correlations in mesoscopic Y
structures in which one of the branches is coupled to a large
resistor. It is clearly of interest to know to what extent special
geometries are necessary for the observation of positive
correlations in purely normal conductors.

\begin{figure}
\includegraphics[height=4cm]{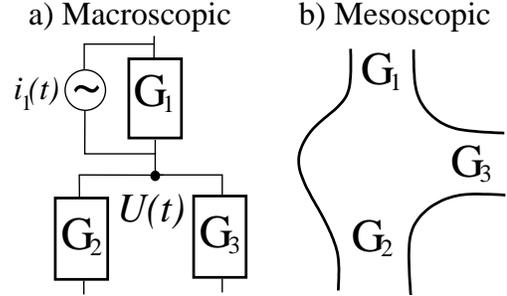}
\caption{
Macroscopic and mesoscopic division of shot noise. The
macroscopic classical circuit of Fig. \ref{fig1}a contains one source of
noise $i_1 (t)$. Fluctuations induced by this source give rise to a positive
current correlation between currents at leads $2$ and $3$. Figure
\ref{fig1}b represents a mesoscopic conductor which  exhibits
negative correlations. $G_i$ are contact conductances. }
\label{fig1}
\end{figure}

In this Letter we point out that under very general conditions there exists
a quantum mesoscopic to macroscopic crossover in purely normal conductors
which manifests itself in the change of sign of current correlations. Figure
\ref{fig1}a depicts a node of a macroscopic conductor in which one
of the branches with conductance $G_1$ contains a source giving rise to shot noise.
The two resistors on the other branches are macroscopic resistors
which generate no shot noise. At the node, the electrostatic potential $U(t)$ must
fluctuate, to ensure conservation of currents. The fluctuating
potential acts in a collective way on electronic carriers,
correlating the currents in the branches. As a consequence such a
classical circuit exhibits positive correlations. In marked
contrast, in the mesoscopic conductor (see Fig. \ref{fig1}b) carriers at different
energies and in different quantum channels are uncorrelated.
Accordingly a carrier exiting through a channel in one of the
leads, leaves an empty state in the other out-going channels, and
as consequence the current correlations are negative.

The
transition from mesoscopic noise division (with negative
current-correlations) to macroscopic noise division (with positive
current correlations) can be investigated in a wide range of
structures. It rests only on the property that correlations
induced by voltage fluctuations can overwhelm correlations due to
the Pauli principle.

{\it Macroscopic versus mesoscopic noise division --}
The classical node (see Fig. \ref{fig1}a) consists of three branches with conductances $G_{\alpha}$.
A voltage $V$ is applied to lead $1$ and the others are grounded.
The conductor $1$ generates shot noise
with a power $p_1 = 2 \int dt \langle
i_1(t) i_1(0) \rangle$. The fluctuating current through this conductor is
$\Delta I_1 (t) =i_1 (t)-G_1 \delta U(t)$.
The conductors $2$ and $3$ are macroscopic and generate
no shot noise. Thus in the zero-temperature limit (which we consider from now on) the
fluctuating current in these branches is $\Delta I_i (t) = - G_i \delta U(t)$ where $U(t)$ is the voltage at the node. From Kirchhoff's law it follows immediately, that the current correlation
$P_{23}=\int dt\langle\Delta I_2(t)\Delta I_3(0)\rangle$  at contacts $2$ and $3$ is
\begin{eqnarray}
\label{classical}
P_{23} =\frac{G_2 \,G_3 }{G_\Sigma ^2} \,p_{1}> 0,
\end{eqnarray}
where $G_{\Sigma} = \sum_i G_i $. In contrast, a mesoscopic quantum
coherent conductor \cite{review,mb91} is described by scattering
matrices $s_{ij} $ which give the current amplitudes in contact $i$ as a function of the incident current amplitudes in contact $j$. With the Fermi distribution in contact $1$ denoted by $f_1$ and the Fermi distributions in contact $2$
and $3$ by $f_0$ the current correlations are \cite{mb91}
\begin{eqnarray}
\label{quantum}
P_{23} =-\frac{2e^2}{h}\int dE \, \Tr[B_{32}^{\dagger}B_{23}](f_1 - f_ 0 )^2\le 0,
\end{eqnarray}
where $B_{23} = s_{21} s^{\dagger}_{13}$ and the trace
is over quantum channels (transverse modes). In fact the correlations
of a quantum coherent conductor are negative  independent of geometry and temperature, number of contacts, etc. The goal of our work is to develop a theory which connects the results of Eq. (\ref{classical}) and Eq. (\ref{quantum}).

{\it Energy conserving transport --} As a generic example, it is instructive to consider
next a chaotic cavity \cite{review1,jala,ober} connected via contacts with conductance
$G_{i}$ to reservoirs. We assume that quasi-elastic scattering is
sufficiently strong such that quantum interference effects are
completely washed out \cite{VLB,bs,PSB}. Since scattering is isotropic, the state of
the cavity can be characterized by the distribution function
$f_c(E)$. This distribution can be found from current conservation
which for quasi-elastic scattering must hold at each energy. The time-averaged
spectral current at contact $i$ is $\bar
I_i(E)=G_i(f_i(E)-f_c(E))$ and from $\sum_i \bar I_i(E)=0$ we find \cite{VLB}:
\begin{equation}
\label{eq:fc_elastic}
 f_c(E) = \frac{G_1 f_1(E)+(G_2+G_3)
f_0(E)}{G_\Sigma}.
\end{equation}

Consider next the fluctuations away from the average current
densities. Contact $i$ with $G_i=(e^2/h)\sum_n T^i_n$, where
$T^i_n$ is the transmission probability of the $n$-th scattering
channel, generates noise \cite{review,mb91,KLes} with a power $p_i
$,
\begin{eqnarray}
\label{eq:noisep} p_i =2 G_i \int dE [f_c(1-f_c)+{\cal
F}_i (f_i - f_c)^{2} ].
\end{eqnarray}
where ${\cal F}_i \equiv  \sum_n T^i_n(1-T^i_n)/\sum_n T^i_n$
is the zero-temperature Fano factor of contact $i$. Here and in the following we have
assumed energy independent transmission probabilities.

The total spectral current fluctuations $\Delta I_i(E,t)$ are composed
of two contributions. The first contribution $i_i(E,t)$ is the current fluctuation
of a conductor with time-independent distribution functions $f_i$ and $f_c$. A second
contribution $G_i\delta f_c(E,t)$ arises from the fact that the distribution function $f_c$
must fluctuate to conserve current at every instant of time.
Thus the total fluctuating spectral current
at contact $i$ is,
\begin{equation}
\label{l1} \Delta I_i(E,t)= i_i(E,t)- G_i\delta f_c(E,t).
\end{equation}
Using the conservation of current fluctuations $\sum_i \Delta
I_i(E,t)=0$ we obtain $\delta f_c = (i_1+i_2+i_3)/G_\Sigma$. Thus
the $\Delta I_i$ can be expressed in terms of the fluctuations of
$i_i$ alone and since the noise sources $i_i$ of different
contacts are independent \cite{bs,PSB} the current
cross-correlations depend only on the auto-correlations, Eq.
(\ref{eq:noisep}). In particular for the current-correlation
$P_{23}=\int dt\langle\Delta I_2(t)\Delta I_3(0)\rangle$  at
contacts $2$ and $3$ we find:
\begin{eqnarray}
\label{eq_corr_elastic} P_{23} =\frac{G_2G_3 \,p_{1} -
G_3(G_1+G_3)\,p_2- G_2(G_1+G_2) \,p_3}{G_\Sigma ^2} .
\end{eqnarray}
We notice that the noise source of contact $1$ gives a positive
contribution see Eq. (\ref{classical}),
whereas the noise sources of contacts $2$ and $3$
contribute with a negative sign. We remark that if all contacts
contain only fully transmitting or fully reflecting modes i.e. by
using a QPC at a plateau they do not produce partition noise
proportional to $G_i{\cal F}_i$. Thus at zero temperature such a cavity
exhibits noise only due to the nonzero "effective temperature" \cite{bs}
$k_B T_{eff}=\int dE f_c(1-f_c)$ inside the cavity. The resulting
correlation is negative and equal to the ensemble averaged shot noise
of a quantum coherent cavity \cite{VLB}.

To reverse the sign of the current correlation one
needs to reduce the negative contribution to Eq. (\ref{eq_corr_elastic})
while keeping the positive contribution
finite.

\emph{Effect of inelastic scattering --} To reduce the negative contributions to the correlations in Eq.
(\ref{eq_corr_elastic}), we now drive
the distribution function $f_c$ inside the cavity toward an equilibrium
distribution function, thereby reducing the
"effective temperature" $k_B T_{eff}=\int dE f_c(1-f_c)$.
We allow particles in the
dot to exchange energy with a voltage probe which is connected to the
cavity with conductance $G_p $.
In the following we consider contact $1$ to generate shot noise with the Fano factor ${\cal F}_1$, and
contacts $2$ and
$3$ to be perfect, so that ${G_2\cal F}_2=0$ and ${G_3\cal
F}_3=0$.

The voltage at the probe is found from
%
$ \bar I_{p}=G_{p}\int dE (f_{p}(E)-f_c(E))=0$,
%
where $G_{p}= (e^2/h) N_{p}$ is the conductance of the contact which
connects the voltage probe with the cavity. Since the distribution
function in the cavity is defined from the balance of currents at
each energy, it will be affected by the presence of the voltage
probe,
\begin{equation}
\label{eq:fc_inel}
 f_c(E)=\frac{G_1 f_1(E)+ G_{0} f_0(E)+G_{p}
f_{p}(E)}{G_\Sigma +G_{p}}.
\end{equation}
Here, as above, $G_\Sigma = \sum_{i=1}^{3} G_i$ and $G_0 \equiv
G_2 + G_3$. Substituting Eq. (\ref{eq:fc_inel}) into
the equation for the current to the probe and performing the integration over energy
we find the voltage in the probe $V_{p}=V_1G_1/G_\Sigma$.
The distribution $f_c $ in the cavity
has three steps (in the elastic case, Eq. (\ref{eq:fc_elastic}),
it has only two steps). If $G_{p}$ is small (weak inelastic
scattering), then the distribution function in the cavity
coincides with the elastic one and strongly deviates from
an equilibrium distribution. In the limit of strong energy relaxation $f_c = f_p $
is an equilibrium distribution with the Fermi energy at $eV_{p}$. The current at the
voltage probe fluctuates according to
\begin{eqnarray}
\Delta I_{p}(E,t)&=&i_{p}(E,t)+G_{p}(\delta f_{p}(E,t) - \delta
f_c(E,t)),
\end{eqnarray}
which together with Eqs. (\ref{eq:noisep},\ref{l1}) fully
specifies the fluctuating currents.

From $\Delta I_{p}(t)= 0$ we obtain $\Delta I_i(t)= i_i (t) - G_i
(i_1 (t) +i_2 (t) +i_3 (t))/G_\Sigma$ where $i_i (t) = \int dE
\,i_i (E,t)$. Therefore the calculation proceeds as above and we
obtain the current cross-correlations at contacts $2$ and $3$:
\begin{eqnarray}
\label{eq:in_positive} P_{23}=\frac{-2G_2G_3}{G_\Sigma }\int
dE [(f_c(1-f_c)-\frac{G_1{\cal F}_1}{G_\Sigma }(f_1-f_c)^2].
\end{eqnarray}
Using expression (\ref{eq:fc_inel}) for the distribution function
and performing the integration over energy we obtain
\begin{eqnarray}
\label{most} P_{23}&=&\frac{-2eV G_1 G_2G_3}{G_\Sigma (G_\Sigma
+G_{p})^2}\left[G_{0}+G_p+\frac{G_p(
G_{0}-G_1)}{G_\Sigma}\right.\nonumber\\
&-&\left.{\cal F}_1\left(\frac{(G_{0}+G_p)^2}{G_\Sigma}-
\frac{G_1G_p(2G_{0}+G_p)}{G_\Sigma^2}\right)\right].
\end{eqnarray}
For $G_{p}=0$ we find the negative
result for the cross-correlations of Eq. (\ref{eq_corr_elastic}), while
for strong inelastic scattering $G_{p}\rightarrow\infty$
we obtain
\begin{equation}
P_{23}=2eV {\cal F}_1 \frac{G_1G_2G_3 G_{0}}{G_\Sigma^3}.
\end{equation}
We see that cross-correlations are indeed positive in the case of
strong inelastic scattering inside the dot.

\begin{figure}
\includegraphics[height=5.0cm]{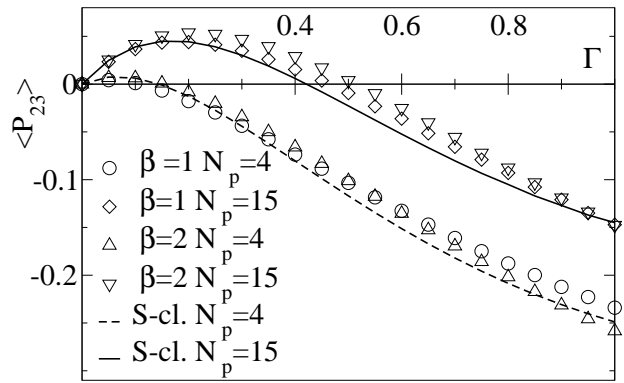}
\caption{Ensemble averaged current-current correlation as a
function of transmission probability $\Gamma$. The full and broken lines are the analytical
results of the semi-classical theory. The open symbols are from a
numerical integration over an ensemble of random matrices for
different symmetry classes $\beta = 1, 2$ and different number of
channels $N_p$ of the voltage probe.} \label{fig_av}
\end{figure}

Eq. (\ref{most}) is a key result of this work. The crossover from
negative to positive cross correlations of Eq. (\ref{most}) are
depicted in Fig. \ref{fig_av} for the case that all
contacts have two channels. The transmission probabilities of the
noise generating contact are both equal and given by $\Gamma$. The
broken line and the solid line are for voltage probes with $4$ and
$15$ channels. For small $\Gamma$ the distribution $f_c$ is very close to an
equilibrium distribution function, and a small amount of inelastic
scattering is sufficient to equilibrate the distribution. As a
consequence for small $\Gamma$ the correlations are positive. As
the transparency increases, the distribution function $f_c$
deviates strongly from the equilibrium Fermi function. The cavity
is effectively "hot" and eventually the cooling provided by
the voltage probe is not sufficiently strong to suppress the
negative contributions to the shot noise correlation. Comparison
of the curves for $N_p =4$ and $N_p = 15$ shows that the range of
positive correlations is the wider the stronger the cooling of the
voltage probe.

\emph{Random Matrix Theory --} We next discuss the crossover for samples which are at least partially coherent.
 Since each sample has particular scattering
properties depending on the shape of the cavity, positions of
impurities, gate voltages it is interesting to consider the
statistical distribution $P$ of the shot noise correlation $P_{23}$.
We describe the node with a scattering matrix and as above introduce
inelastic scattering with a voltage probe.
For chaotic cavities, the case of interest here, the ensemble
averaged noise is equal to the semi-classical result Eq.
(\ref{most}). In the presence of fluctuations away from the
ensemble average there is therefore the interesting possibility
that ensemble members might have correlations with a different sign than the ensemble
averaged correlation.

Statistical properties of the transport quantities of chaotic
cavities are well described by Random Matrix Theory (RMT). From
our discussion we know that additional scattering at the contact
$1$ is essential in order to reverse sign of cross-correlations.
If the quantum channels of the contacts are not fully transparent
then the combined scattering matrix of the cavity and contacts can
be written \cite{BrBee}:
$S=\hat{R}-\hat{T}(\hat{1}-\hat{U}\hat{R})^{-1}\hat{U}\hat{T}$.
Here $\hat{R}$ and $\hat{T}$ are reflection and transmission
matrices of contacts.
We chose scattering determined by $\hat\Gamma$ at a contact $1$
and perfectly transmitting channels in all other contacts.
Therefore $\hat{T}=diag\{\sqrt{\hat \Gamma},1,1...\}$ and
$\hat{R}=diag\{\sqrt{1-\hat \Gamma},0,0..\}$ where $\hat\Gamma$ is
the transmission matrix of the contract $1$. $U$ is scattering
matrix of the cavity itself which is distributed uniformly over
the orthogonal, $\beta=1$, (unitary, $\beta=2$) ensemble. For
simplicity we keep all transmission probabilities equal and take
$\hat \Gamma_n=\Gamma$.

\begin{figure}[bottom]
\includegraphics[height=5cm]{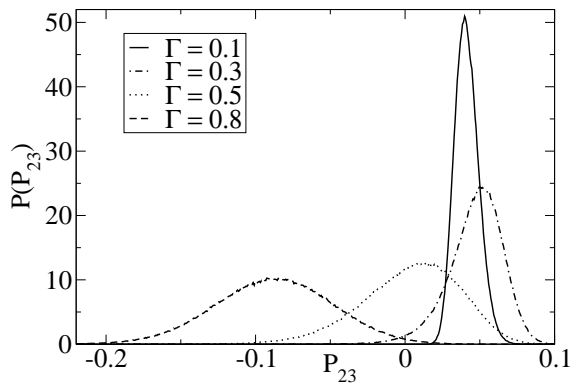}
\caption{Distribution of the shot noise correlation for a
mesoscopic cavity. The cavity is connected via a noise generating
contact with transmission $\Gamma$ per channel and two perfect
contacts to reservoirs. For the case of two channels per contact
and 15 channels in the voltage probe the transition from negative
to positive correlations is near $\Gamma = .5$.}
\label{fig_dist}
\end{figure}

In the presence of a voltage probe the cross correlation
$P_{23}$ can be expressed in terms of the noise correlators
$p_{ij}= 2\int dt \langle i_i(t)i_j(0)\rangle$ calculated for a
conductor with equilibrium distribution $f_1,f_p,f_0$ at the
corresponding contacts. The correlation $P_{23}$ is given
\cite{BNazarov,texier,VLB} by
\begin{eqnarray}
\label{eq:RMT_cross_corr}
P_{23}=p_{23}-\frac{G_{2p}p_{3p}}{G_{pp}}
-\frac{G_{3p}p_{2p}}{G_{pp}}+\frac{G_{2p}G_{3p}p_{pp}}{G_{pp}^2}.
\end{eqnarray}
Here
$G_{\alpha\beta}=(e^2/h)[\delta_{\alpha\beta}N_\alpha-\Tr(S^\dag_{\beta\alpha}S_{\alpha\beta})]$
are the coherent conductances determined from the
scattering matrix $S_{\alpha\beta}$.

The results of a numerical integration of Eq.
({\ref{eq:RMT_cross_corr}) for the ensemble averaged cross
correlation are shown in Fig. \ref{fig_av} for $\beta =1,2$ and
$N_p =4,15$ and compared with the semiclassical result Eq.
(\ref{most}). For $\Gamma =1$ it is straightforward to evaluate
\cite{BrBee} Eq. (\ref{eq:RMT_cross_corr}) to leading order in the
number of channels for $\beta=2$,
\begin{equation}
\label{eq:RMT_analytics}
P_{23}=-\frac{4e^3V}{9h}\frac{N_L^2(3N_L+2N_{p})}{(3N_L+N_{p})^2},
\end{equation}
where $N_L$ is the number of channels in each lead, $N_p$ is the number of
channels in the probe. There
is a perfect agreement between the semiclassical result of Eq.
(\ref{most}), numerical integration of Eq.
(\ref{eq:RMT_cross_corr}) for $\Gamma=1$, and the analytical
calculation Eq. (\ref{eq:RMT_analytics}).

We now obtain numerically the full statistical distribution $P$ of the
current cross-correlations Eq. (\ref{eq:RMT_cross_corr}). Figure
\ref{fig_dist} shows a set of distribution functions of current
cross-correlations $P_{23}$ for different transparency $\Gamma$ of
the contact $1$ for $\beta=2$. For very small $\Gamma$ the
distribution function is large only for positive values of the
correlation. As $\Gamma$ becomes larger a tail of the distribution
extends to the region of negative correlations. Eventually for
large $\Gamma$ the distribution is large only for negative
correlations with a tail extending to positive values of $P_{23}$.

{\it Conclusions --}
Photons bunch, electrons anti-bunch! This statement is often made to explain the negative sign of current correlations in mesoscopic conductors. However, electrons are interacting entities. In particular, voltage fluctuations which accompany inelastic scattering introduce correlations which are stronger than those dictated by the Pauli principle alone. As a consequence, under a wide range of conditions, current-current correlations in normal mesoscopic conductors can be positive. We examined the crossover in detail for a range of geometries that can be subjected to experimental tests. Importantly our work demonstrates that positive correlations can in general not be used as  an 'entanglement witness' since they can be due to purely classical correlations.

We thank M. Polianski, S. Oberholzer, and E.V. Sukhorukov for discussions.
This work is supported by the Swiss NSF.

\end {document}